\documentclass[pra, showpacs, preprint, superscriptaddress]{revtex4}
\usepackage{longtable}
\usepackage{graphicx}    % Include figure files
\usepackage{bm}          % bold math%
\usepackage{dcolumn}    % Align table columns on decimal point
\usepackage{amssymb}
\usepackage{color}
\usepackage{CJK}
\begin{document}
%\begin{CJK*}{GBK}{song}
\title{A Frequency Filter of Backscattered Light of Stimulated Raman Scattering due to the Raman Rescattering in the Gas-filled Hohlraums}

\author{Liang Hao}
\affiliation{Institute of Applied Physics and
Computational Mathematics, Beijing, 100094, China}

\author{Wenyi Huo\footnote{Email: huo\_wenyi@iapcm.ac.cn}}
\affiliation{Institute of Applied Physics and Computational
Mathematics, Beijing, 100094, China}

\author{Zhanjun Liu\footnote{Email: liuzj@iapcm.ac.cn}}
\affiliation{Institute of Applied Physics and Computational
Mathematics, Beijing, 100094, China}

\author{Chunyang Zheng}
\affiliation{Institute of Applied Physics and Computational
Mathematics, Beijing, 100094, China}

\author{Chuang Ren}
\affiliation{Department of Mechanical Engineering and Laboratory for
Laser Energetics, University of Rochester, Rochester, New York
14627, USA} \affiliation{Department of Physics and Astronomy,
University of Rochester, Rochester, New York 14627, USA}

%\author{Jun Li} \affiliation{University of California at San Diego, USA}

%\author{C. Ren\footnote{Email: chuang.ren@rochester.edu}}
%\author{C. Ren\footnotetext[2]{Email:cren2@ur.rochester.edu}}
%\affiliation{Department of Mechanical Engineering and Laboratory for Laser Energetics, University of Rochester,
%Rochester, New York 14627, USA}
%\affiliation{Department of Physics and Astronomy, University of Rochester, Rochester, New York 14627, USA}

\date{\today}% It is always \today, today, %  but any date may be explicitly specified

\begin{abstract}
The coupling evolutions of stimulated Raman scattering (SRS) and
Raman rescattering (re-SRS) are studied under the parameter
conditions of relevance to the gas-filled hohlraum experiments at the National Ignition Facility by a
nonenveloped fluid code for the first time. It is found that re-SRS
works as a frequency filter of backscattered light of SRS in the gas
region. The low frequency modes originated from density points
higher than about $0.1n_c$ would stimulate re-SRS and be heavily
depleted by re-SRS at the region of their effective quarter critical
density region. Due to the high collisional damping of the
rescattered light, the energy of rescattered light is deposited
quickly into the plasmas along with its propagation, which limits
the re-SRS in a small region. Large amplitude of the daughter
Langmuir wave of re-SRS would stimulate cascade Langmuir decay
instabilities and induce obvious low frequency density modulations,
which can further result in the inflation of high frequency modes
generated at density points lower than the growth region of re-SRS.

\end{abstract}

\pacs{52.50Gi, 52.65.Rr, 52.38.Kd}% PACS, the Physics and Astronomy
\maketitle

%\section{Introduction}

Laser plasma instabilities (LPIs) \cite{Kruer1988}, such as
stimulated Raman scattering (SRS), stimulated Brillouin scattering
(SBS), are of great importance in laser driven inertial confinement
fusions. In direct drive inertial confinement fusion, SRS and SBS
can reduce the laser absorption. SRS would generate energetic
electrons which may preheat the fuel. SBS would result in cross beam
energy transfer (CBET) between overlapped laser beams and drive
asymmetry \cite{Craxton2015,Campbell2017}. In indirect drive
inertial confinement fusion, SRS and SBS would scatter the laser
energy out of hohlraums which are used to convert laser energy into
soft X rays. As a result, SRS and SBS lead to the reduction of the
laser energy absorbed by the hohlraum wall and the conversion
efficiency from laser into soft X rays. The CBET would redistribute
the laser deposition and result in drive asymmetry
\cite{Lindl1995,Lindl2014,Moody2012}. Besides, SRS and two plasmon
decay are important sources of energetic electrons in both direct
and indirect drives \cite{Myatt2014}.

Generally, indirect drive inertial confinement fusion uses
gas-filled hohlraums to mitigate the flow of wall and ablator
plasmas into the interior of the hohlraum
\cite{Moody2014,Lanke2016,Huo2018}. In the past years, there have
been many works engaged in the study of LPIs in the gas-filled
hohlraums
\cite{Kirkwood2013,Hall2017,Strozzi2017,Hinkel2008,Hinkel2011,Berger2019,Michel2012}.
These works mainly concentrate on the primary LPIs mentioned above,
based on the linear ray-racing model \cite{Hall2017,Strozzi2017},
enveloped fluid model \cite{Hinkel2008,Hinkel2011,Berger2019}, or
kinetic simulation \cite{Michel2012}. However, the experimental
results at the National Ignition Facility (NIF) indicate that the
LPIs in the gas-filled hohlraums are too complicated and difficult
to be well understood. For example, the SRS level of the laser beam
at 30$^\circ$ seems to be saturated and do not go up with the
transferred energy from outer cones by CBET \cite{Michel2009}, and
the wavelength of SRS backscattered light is commonly shorter than
simulations of the linear ray-tracing code
\cite{Hall2017,Strozzi2017}. A lack of understanding and controlling
LPIs is believed to be one of major contributors to the failure to
achieve ignition \cite{Hinkel2016}.

In fact, besides the primary LPIs, some secondary LPIs, such as the
Langmuir decay instability (LDI) \cite{DuBois1965}, and the
rescattering of the scattered light of the primary LPIs, including
Raman rescattering (re-SRS) \cite{Hinkel2004}, Brillouin
rescattering \cite{Montes1985}, and the two plasmon decay
\cite{Pan2018}, may also play an important role in the gas-filled
hohlraums. Using particle-in-cell simulations, Hinkel {\it et al.}
pointed out that the rescattering could appear in the NIF hohlraums
\cite{Hinkel2004}, and the rescattering of SRS was found to be one
of the important sources of energetic electrons with energy higher
than 100 keV \cite{Winjum2013}. In this Letter, we studied the
coupling evolution of SRS and re-SRS with nonenveloped fluid code in
the indirect drive regime for the first time. It is found that the
re-SRS, which is not only one of the nonlinear saturation mechanism
of SRS stimulated from the high density region, but also can induce
the inflation of SRS generated in the low density region, can work
as a frequency filter of backscattered light of SRS. The results
explain why the observed light wavelength of SRS is shorter than the
linear simulations for the inner laser cones in the NIF gas-filled
hohlraum experiment. In addition, our study can partially explain
the ``energy deficit" appeared in the gas-filled hohlraum
experiments at the NIF \cite{MacLaren2014}.

%\section{Fluid simulations}

%This condition corresponds to the plasma status that the electron
%temperature $T_e$ is lower than $2.5$ keV and the normalize electron
%density $n_e$ is less than $0.11$ $n_c$, where $n_c$ is the critical
%electron density.

Commonly, in gas-filled hohlraum experiments at the NIF, there is a
large inhomogeneous plasma region in gas with the maximum electron
density higher than $0.1 n_c$, where $n_c$ is the critical density.
Theoretically, the scattered light of convective backward SRS (BSRS)
can stimulate re-SRS below the density points $0.11 n_c$. In order
to explore the possible coupling between BSRS and re-SRS in
experiments, in this study, we mainly focus on the inhomogeneous
linear density profile ranged from 0.08 $n_c$ to 0.11 $n_c$ with a
large scale length of about $2$ mm, which is similar to the plasma
conditions in the NIF gas-filled hohlraums
\cite{Lindl2004,Hall2017}. The pump laser wavelength $\lambda_0$ is
0.351 $\mu$m, and ion species is ionized Helium, which are typically
used in ICF experiments. Temperatures are $T_e$ = 1.5 keV for
electrons and $T_i$ = 1 keV for ions. So the production of the
wavenumber of Langmuir wave (LW) $k_L$ and the Debye length
$\lambda_D$ maintained smaller than 0.3 for both SRS and re-SRS,
where the kinetic effects are not important in the fluid regime. Two
one-dimensional simulations with fixed and mobile ions were done
with FLAME code, which is a fluid code based on the full wave
equations for vector potentials of light without any envelope
approximations and can cover the coupling evolution of SRS, SBS,
LDI, and rescattering inherently in the inhomogeneous plasma
\cite{Hao2017}. Seed level can be controlled in this code, which
avoids the impacts of high artificial noises on LPIs which often
appears in Particle-in-Cell simulations \cite{Okuda1972,Hao2016b}.
Currently, in order to ensure the reasonable collisional damping
rates for the light signals with different frequencies, we have
improved the code by substituting the quiver velocity
$v_{\alpha,\beta}$ for the vector potential $A_{\alpha,\beta}$ in
the right hand side coupling terms of the electron and ion momentum
equations and replacing the original equations of vector potential
by $\left(
\partial ^2/\partial t^2 -
\partial ^2/\partial x^2\right)A_\alpha = - n_L v_\beta-n_A v_\alpha$
combined with $\left( {\partial /\partial t} + \nu _{ei}
\right)v_\alpha =\partial A_\alpha/\partial t$, where subscripts
$\alpha, \beta =0$ or $1$, and $\nu_{ei}$ is the electron-ion
collisional damping rate \cite{Hao2017}. In the simulations, the
length of entire simulation box is 16800 $c/\omega_0$, including the
physical domain $L$ = 12000 $c/\omega_0$ (about 670 $\mu$m) and two
perfect matched layers \cite{Berenger1994} of $L_{lay}$ = 2400
$c/\omega_0$ each at the left and right boundaries, which are used
to ensure a good absorption of signals out of the physical
boundaries. The magnitude of volume noises is $10^{-9}$ $n_c$ in
order to describe the thermal noises \cite{Berger1998}. And each
simulation is ran about 40 ps with laser intensity $I_0$ = 2
$\times$ $10^{15}$ W/cm$^2$.

%\subsection{Coupling of SRS and re-SRS with fixed ions}

\begin{figure}[htb!]
\includegraphics[height=0.15\textwidth,width=0.23\textwidth,angle=0]{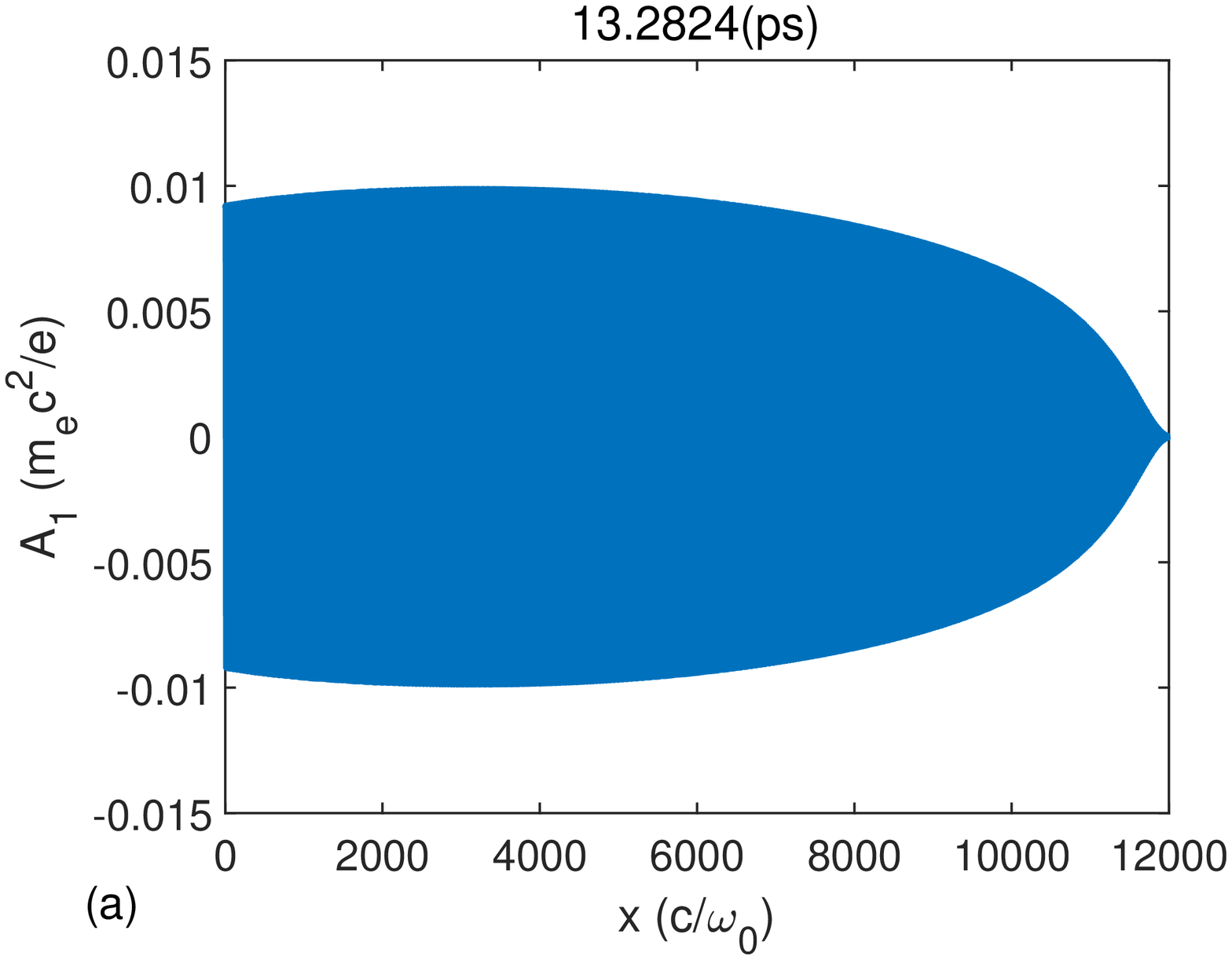}
\includegraphics[height=0.15\textwidth,width=0.23\textwidth,angle=0]{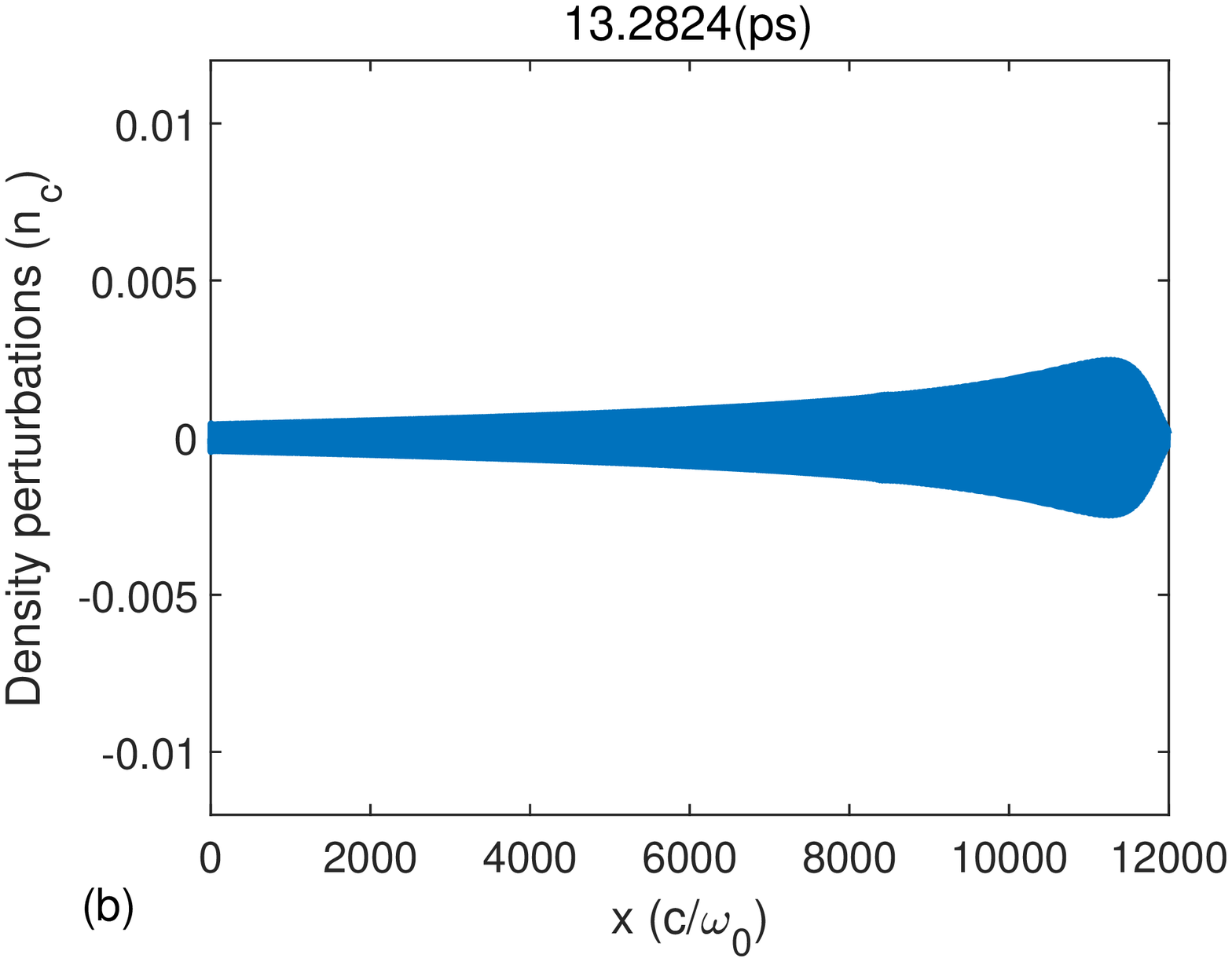}
\\
\includegraphics[height=0.15\textwidth,width=0.23\textwidth,angle=0]{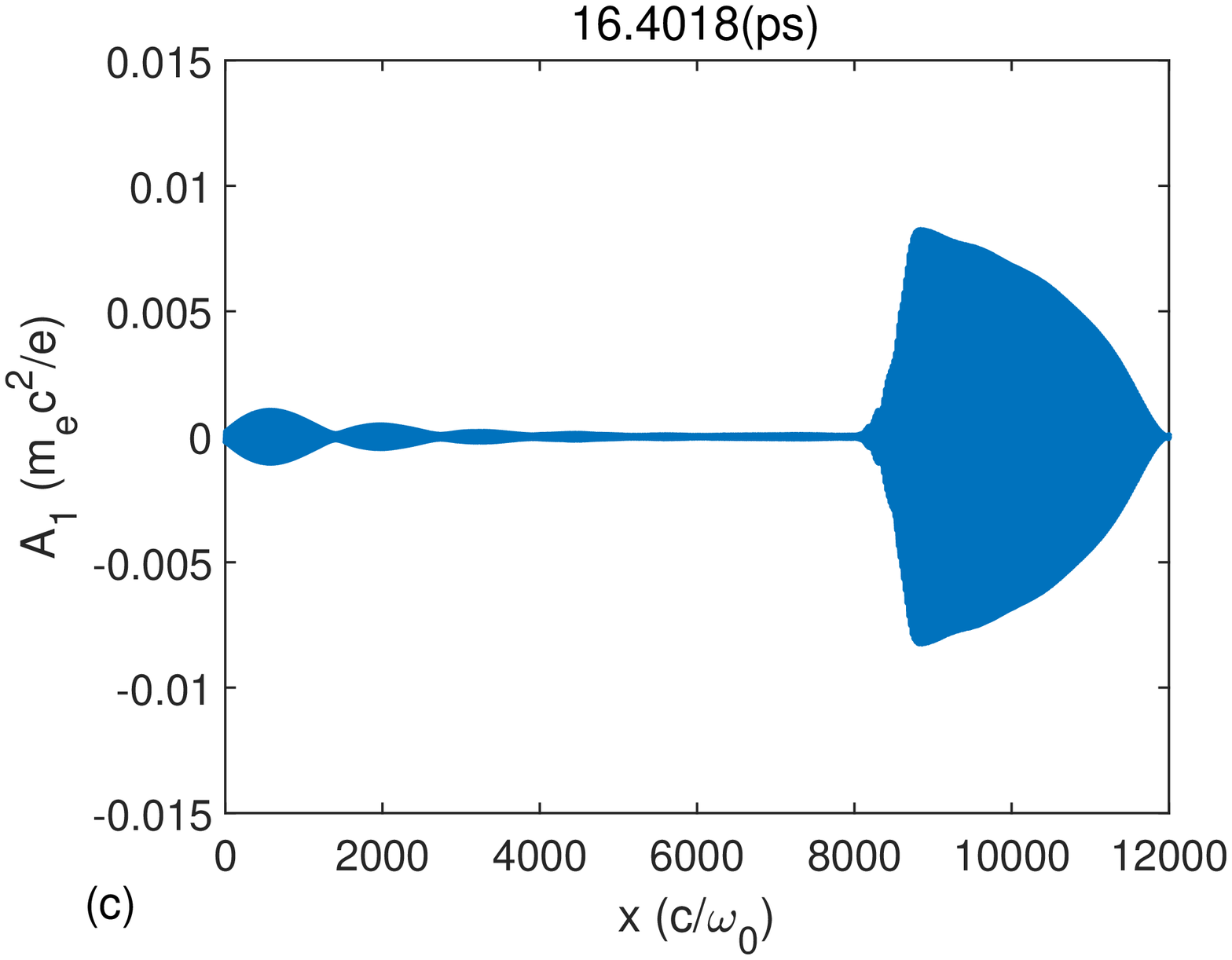}
\includegraphics[height=0.15\textwidth,width=0.23\textwidth,angle=0]{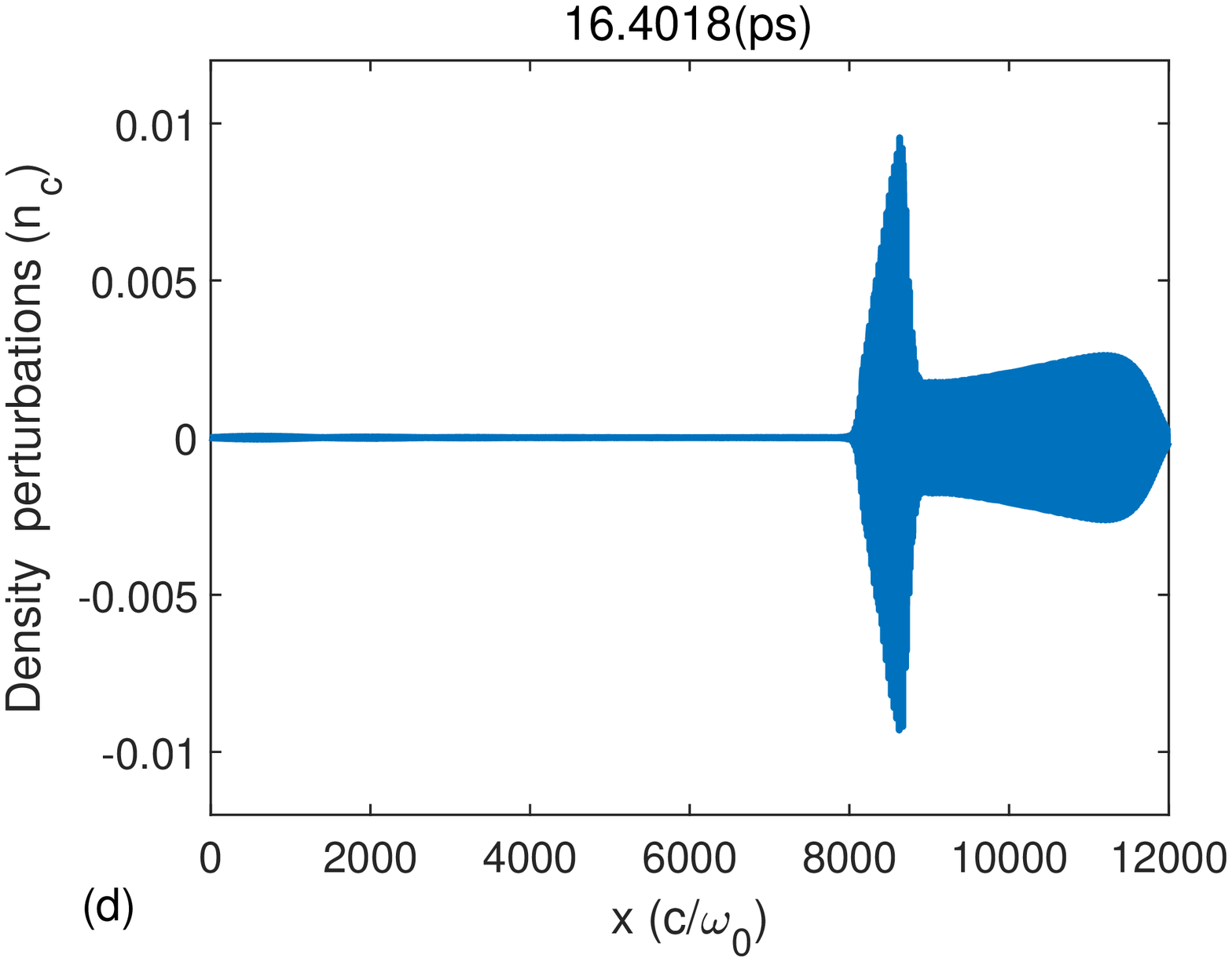}
\caption{(Color online) Snapshots of (a,c) vector potential of SRS
scattered light and (b,d) high frequency density perturbations due
to the LWs of SRS and re-SRS at about (a,b) $13.3$ps and (c,d)
$16.4$ps for the fixed ions case.}
\label{fig_snapshots}
\end{figure}

\begin{figure}[htb!]
\includegraphics[height=0.15\textwidth,width=0.23\textwidth,angle=0]{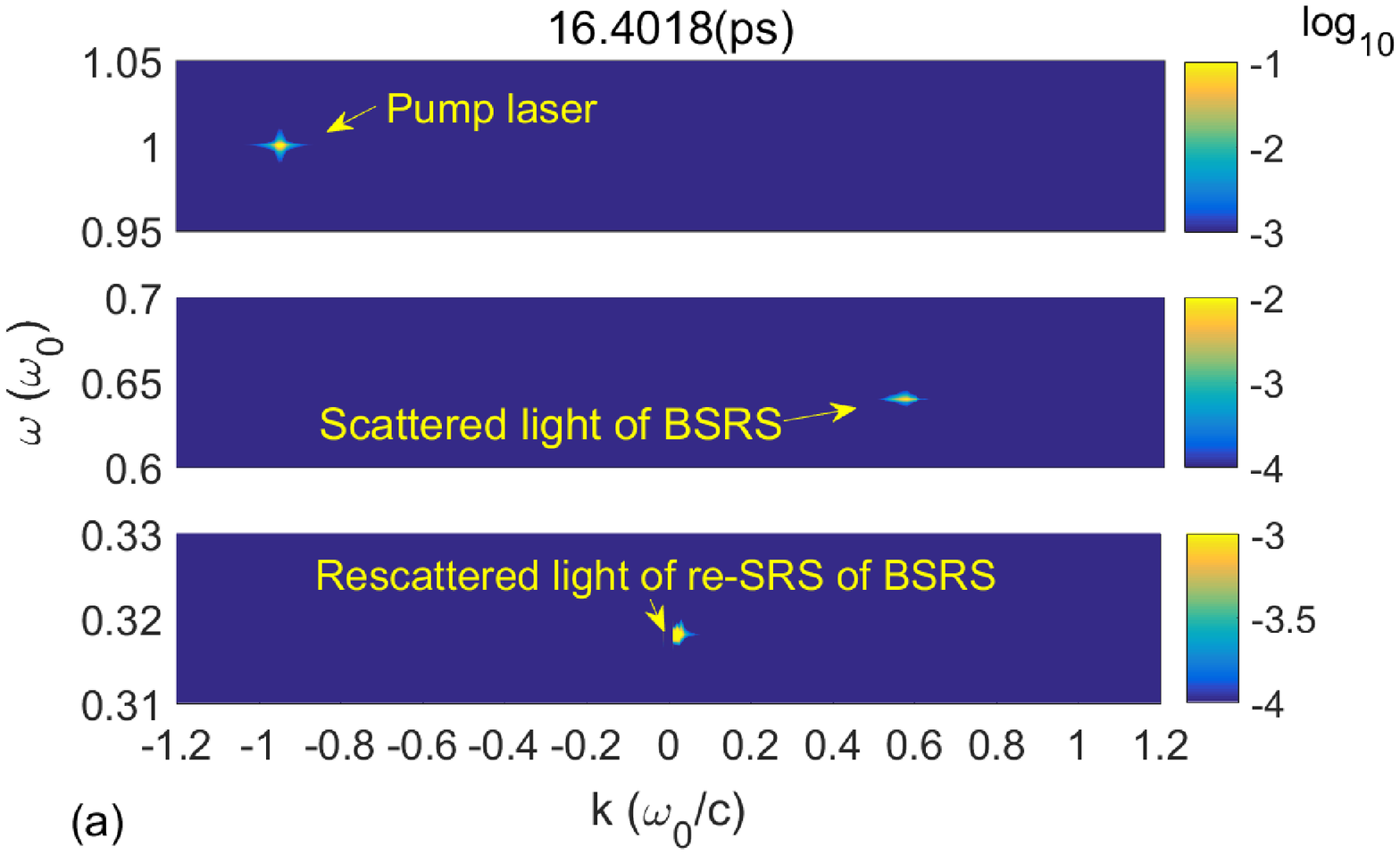}
\includegraphics[height=0.15\textwidth,width=0.23\textwidth,angle=0]{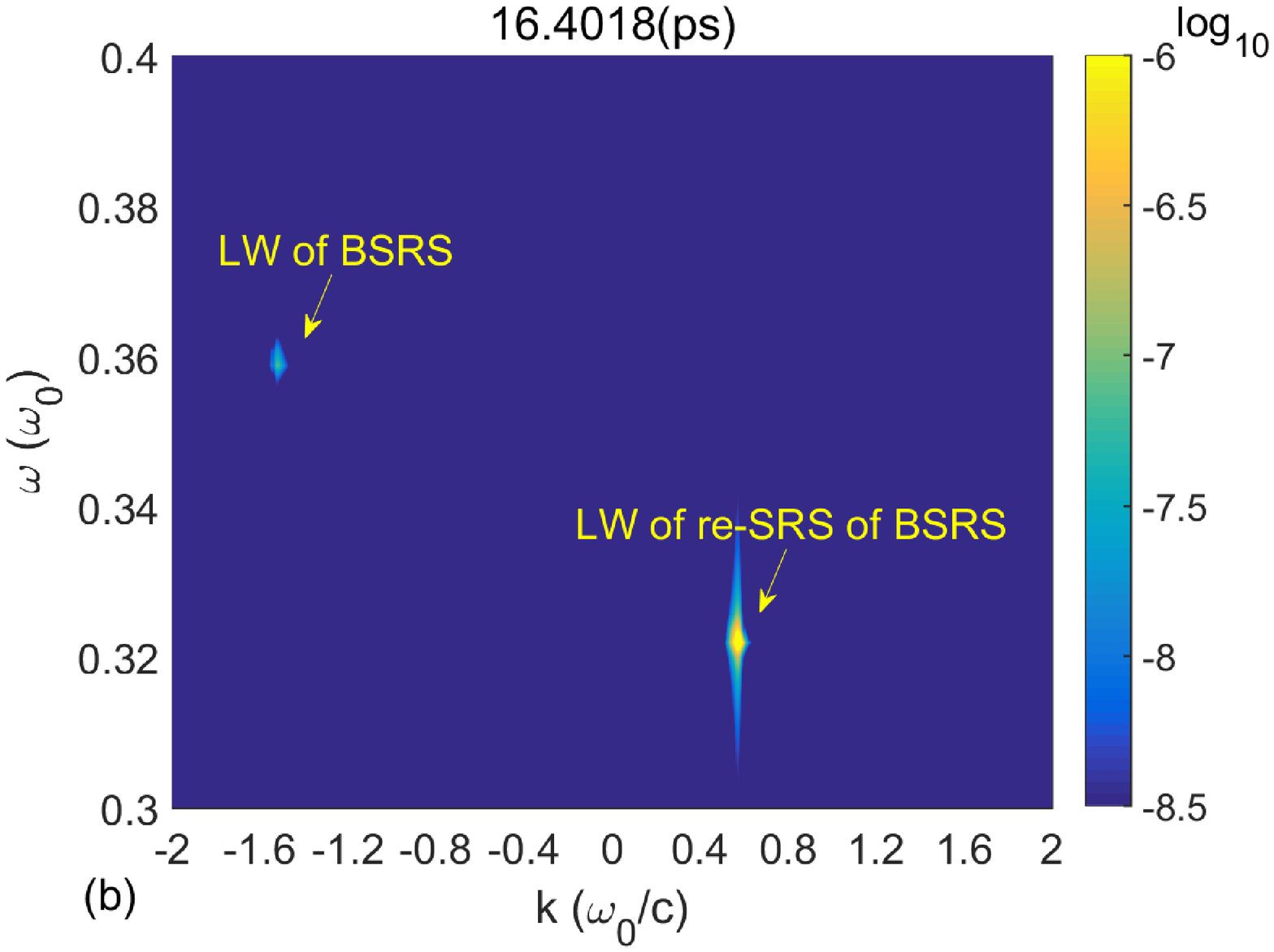}
\caption{(Color online) $\omega-k$ space of (a) light wave and (b)
LWs of BSRS and re-SRS in the range of $8200$-$8800c/\omega_0$ at
the time about $16.4$ps for the fixed ions case.}
\label{fig_psnapshots}
\end{figure}

In order to show the process of re-SRS clearly, we first investigate
the case with fixed ions. The vector potential of SRS scattered
light $A_1$ and the high frequency electron density perturbations of
LWs are shown in Fig. \ref{fig_snapshots}. The spectra of light
waves and LWs in $\omega-k$ space are shown in
Fig.\ref{fig_psnapshots}, which are obtained in the region of
$8200$-$8800c/\omega_0$ (the density is $0.1005$-$0.102n_c$) by
using the 2-dimensional Fourier transform. Signals on the left half
side with ($k<0$) and on the right half side with ($k>0$) represent
the waves propagating forward to the right and propagating backward
to the left in real space, respectively.

\begin{figure}[htb!]
%(a)
%\includegraphics[height=0.35\textwidth,width=0.45\textwidth,angle=0]{T_vs_t_compare.eps}
%(b)
\includegraphics[height=0.3\textwidth,width=0.4\textwidth,angle=0]{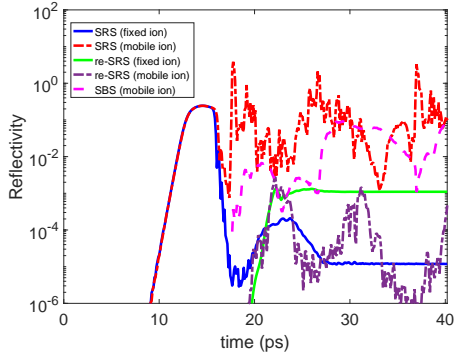}
\caption{(Color online)
%(a) Transmittance of laser intensity at the right boundary and (b)
Reflectivity of backward scattered light at the left boundary versus
time.
%in fluid simulations.
} \label{fig_TR}
\end{figure}

As shown in Figs. \ref{fig_snapshots}(a) and \ref{fig_snapshots}(b),
the convective BSRS rise firstly near the right boundary before the
growth of re-SRS, because the temporal growth rate of BSRS is
commonly larger at higher density. Scattered light of BSRS with the
frequency about $\omega_R$ = 0.64 $\omega_0$ propagates from right
to left and is amplified along with its propagation, accompanied
with the growth of LW on its path. Both light wave and LW show the
feature of the single mode in whole physical domain, which indicates
that BSRS is much easier to be stimulated by the seed light
originated from high density region than the local thermal noises in
the low density region. However, the convective amplification of
scattered light became slower in the left side due to the detune
induced by the density gradient. In the region of 8200 - 8800
$c/\omega_0$, the density is just close to the effective quarter
critical density of the scattered light of BSRS, so absolute mode of
re-SRS is easy to be stimulated. As shown in Figs.
\ref{fig_snapshots}(c) and \ref{fig_snapshots}(d), re-SRS rise
clearly in this special region, and the scattered light of BSRS was
strongly depleted by re-SRS. Judging from the spectra shown in Fig.
\ref{fig_psnapshots}, both the rescattered light and the LW of
re-SRS propagates backward to the left side. And re-SRS grew rapidly
to a large level in a small region and then decreased quickly along
the propagation of its daughter waves as shown in Figs.
\ref{fig_snapshots}(d). According to the matching condition of
re-SRS, the frequency and wavenumber of the rescattered light are
0.318 $\omega_0$ and 0.005 $\omega_0/c$ respectively at the
effective quarter critical density of the backscattered light of
BSRS, while the frequency and wavenumber of the backward LW of
re-SRS are 0.322 $\omega_0$ and 0.54 $\omega_0/c$ respectively,
which are consistent with the corresponding modes as shown in Fig.
\ref{fig_psnapshots}. Due to the small $k_L\lambda_D$ of the LW of
re-SRS, which is only about $0.1$ here, Landau damping of LW of
re-SRS is very small. However, the collisional damping of
rescattered light is about 10 times higher than the incident light
because the frequency of rescattered light is very low. So the
collisional damping, which dominated over the Landau damping here,
is the main reason for the threshold of re-SRS and the quick damping
of its daughter waves in space. Another reason for the saturation of
re-SRS is the depletion of the scattered light of BSRS. Fig.
\ref{fig_TR} shows the reflectivity of scattered light and
rescattered light, which is diagnosed in $\omega-k$ space at the
left boundary of physical domain through 2-dimensional Fourier
transform. In the fixed ions case, reflectivity of backscattered
light of BSRS reached its maximum saturated level of about 24.3 $\%$
firstly and then largely decreased from about 16 ps accompanied with
the growth of re-SRS. After the enough growth of re-SRS, SRS and
re-SRS evolves to a steady state finally, and most of the energy in
scattered light of BSRS was transferred into the daughter waves of
re-SRS and then deposited in plasma due to the collisional damping.
Temporal-averaged reflectivity of scattered light of BSRS and
rescattered light of re-SRS were 1.7$\%$ and 0.05$\%$. No forward
SRS was observed because of the density gradient \cite{Winjum2013}.

\begin{figure}[htb!]
\includegraphics[height=0.15\textwidth,width=0.23\textwidth,angle=0]{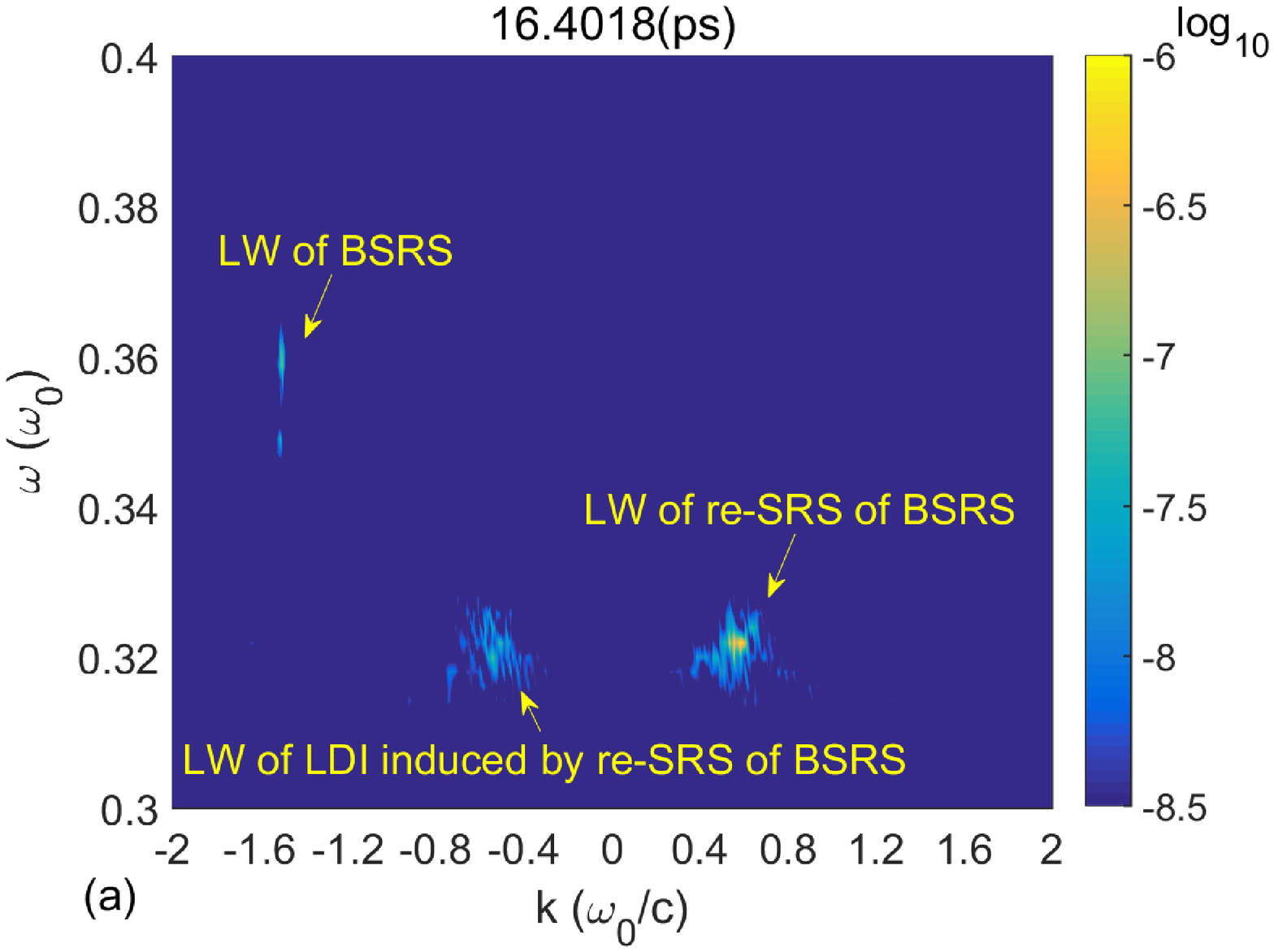}
\includegraphics[height=0.15\textwidth,width=0.23\textwidth,angle=0]{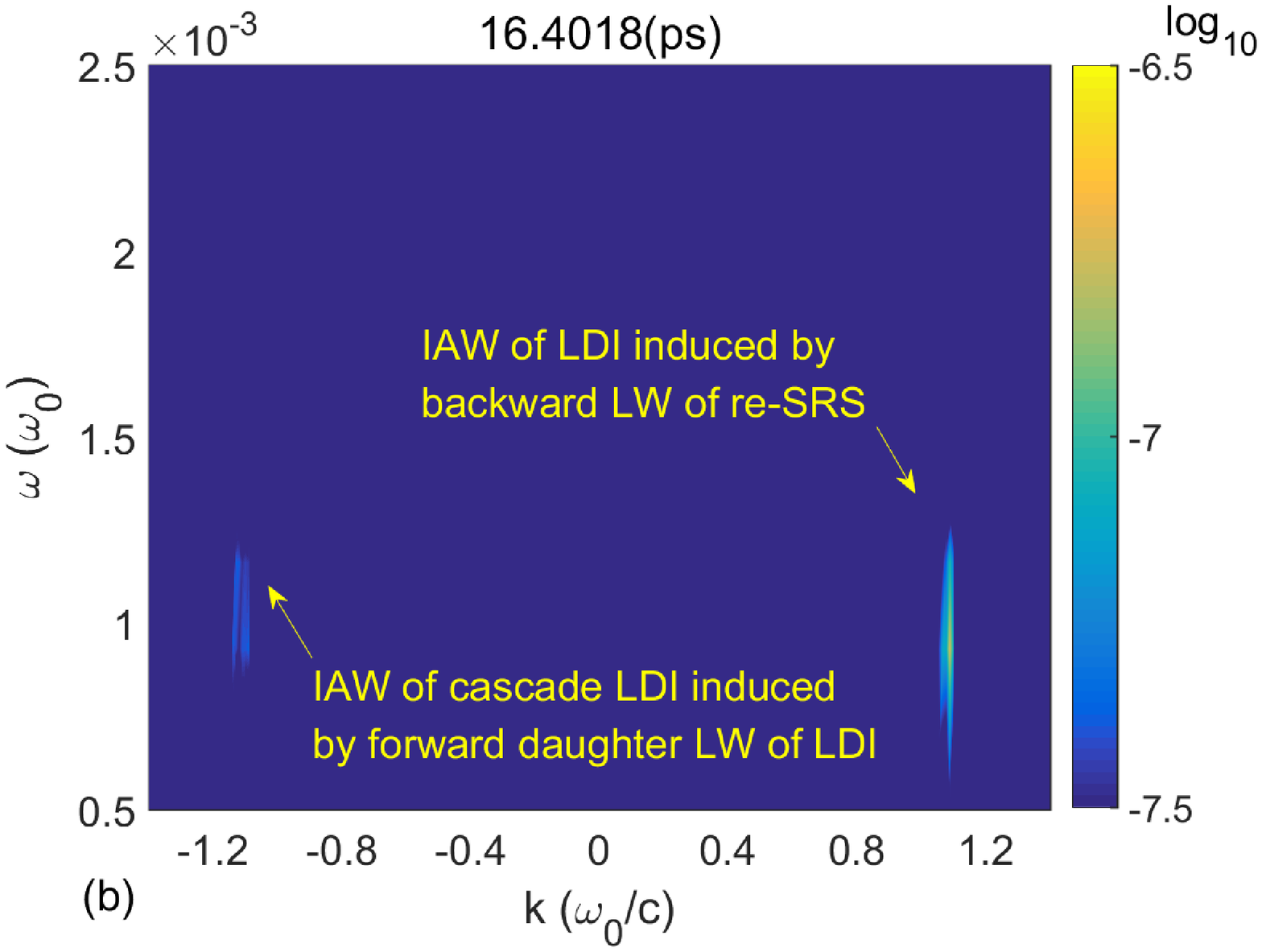}
\caption{(Color online) $\omega-k$ space of (a) LWs of SRS, re-SRS
and LDI, and (b) IAWs of cascade LDI in the range of
$8200$-$8800c/\omega_0$ at the time about $16.4$ps for the mobile
ions case.} \label{fig_ion_psnapshots1}
\end{figure}

In order to be more closer to the real plasma condition in the
gas-filled hohlraum experiments, we considered the case with mobile
ions. As shown in Fig. \ref{fig_TR}, it is interested that the
reflectivity of scattered light of BSRS decreases firstly at 16 ps
and then increases quickly again to a higher saturated level, and
strong bursts are presented. The temporal-averaged reflectivity of
scattered light of BSRS and rescattered light of re-SRS are13.3 $\%$
and 0.01$\%$, respectively. Besides, some observable reflectivity of
scattered light of backward SBS appears from about $17.5$ ps, with
the temporal-averaged value of 1.5 $\%$. The key difference with the
first simulation is that some low frequency density modulations of
ion-acoustic waves (IAWs) occur in the region of 8200 - 8800
$c/\omega_0$, accompanied with the growth of re-SRS. The spectra of
LWs in $\omega-k$ space at the time about 16.4 ps is shown in Fig.
\ref{fig_ion_psnapshots1}(a). In the left half part, besides the LW
of BSRS with frequencies around 0.36 $\omega_0$, there are some new
modes with frequencies around 0.32 $\omega_0$. These modes are the
forward daughter LW of LDI. In order to confirm the existence and
origin of LDI, the spectra of low frequency density modulations in
this region is shown in Fig. \ref{fig_ion_psnapshots1}(b).
Obviously, there are some modes of IAWs with frequency around 0.001
$\omega_0$ and the wavenumber around 1.1 $\omega_0/c$, which is
about two times of the wavenumber of the backward LW of re-SRS but
not the forward LW of BSRS. Therefore, the low frequency density
modulations in this region originates from LDI, which is stimulated
by the backward LW of re-SRS. According to matching condition of
LDI, the modes of IAW in right half side is the backward daughter
IAW of LDI, and the modes in the left half side indicate the cascade
LDI induced by the forward daughter LW of LDI as labeled in Fig.
\ref{fig_ion_psnapshots1}(b).

\begin{figure}[htb!]
\includegraphics[height=0.15\textwidth,width=0.23\textwidth,angle=0]{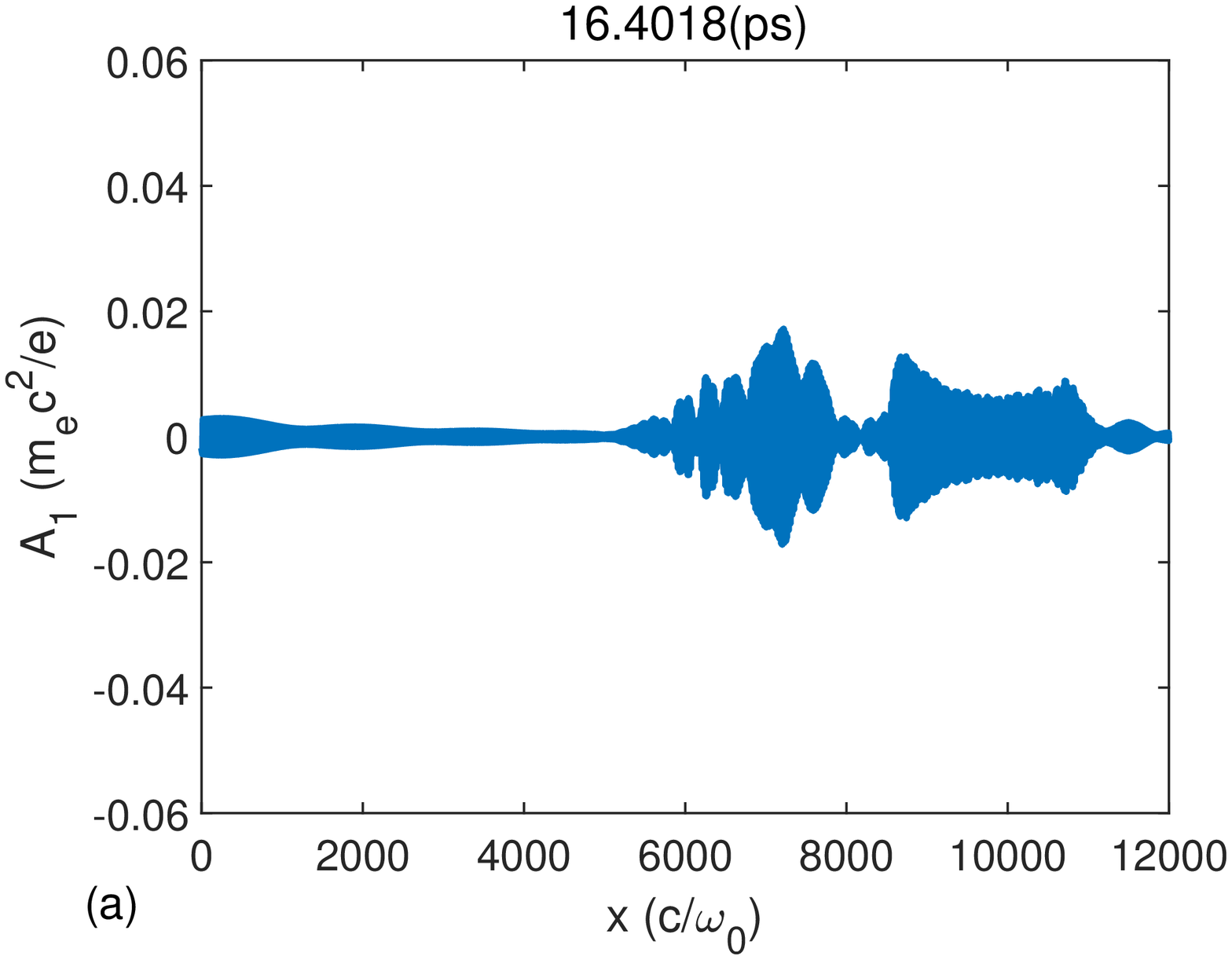}
\includegraphics[height=0.15\textwidth,width=0.23\textwidth,angle=0]{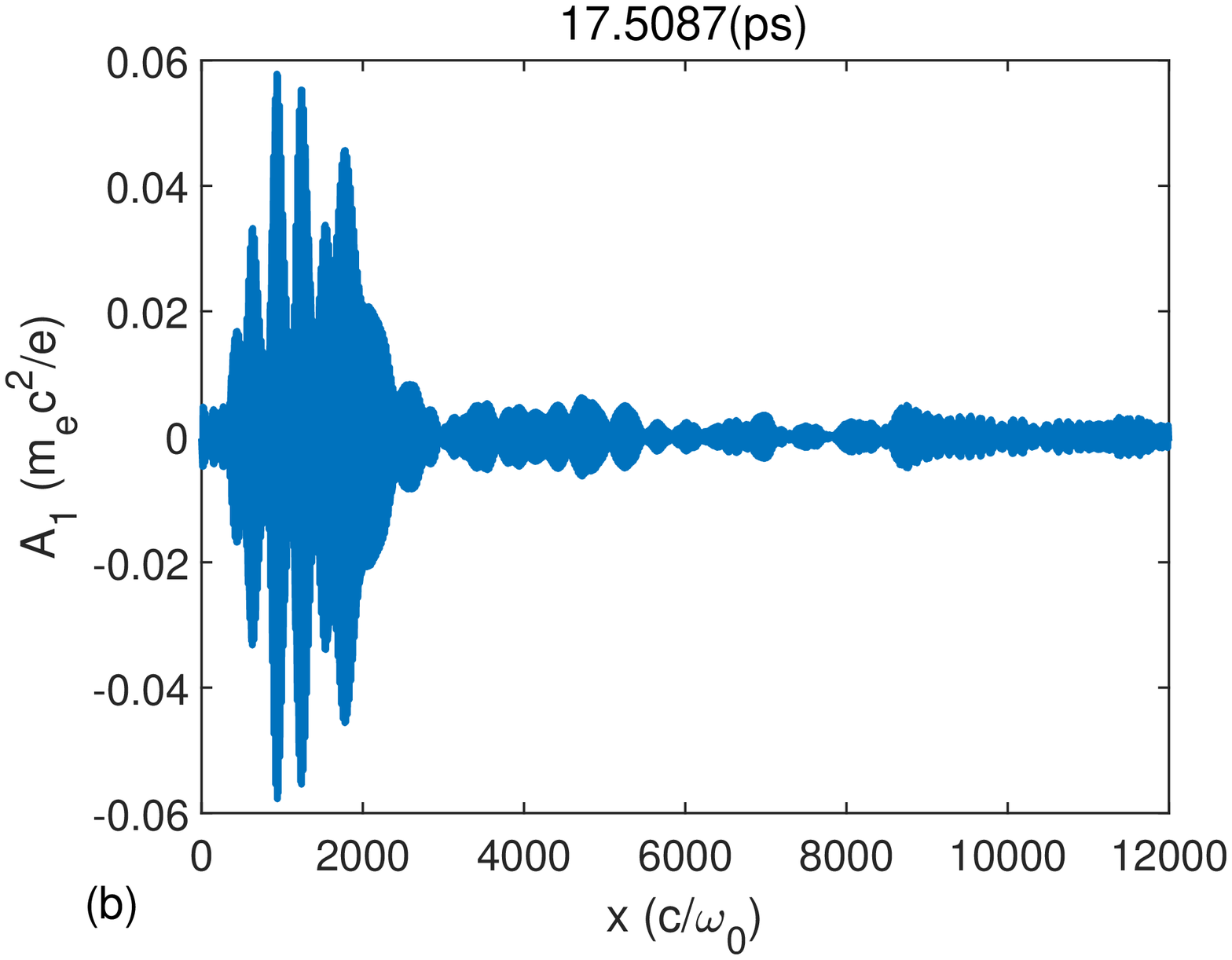}
\includegraphics[height=0.15\textwidth,width=0.23\textwidth,angle=0]{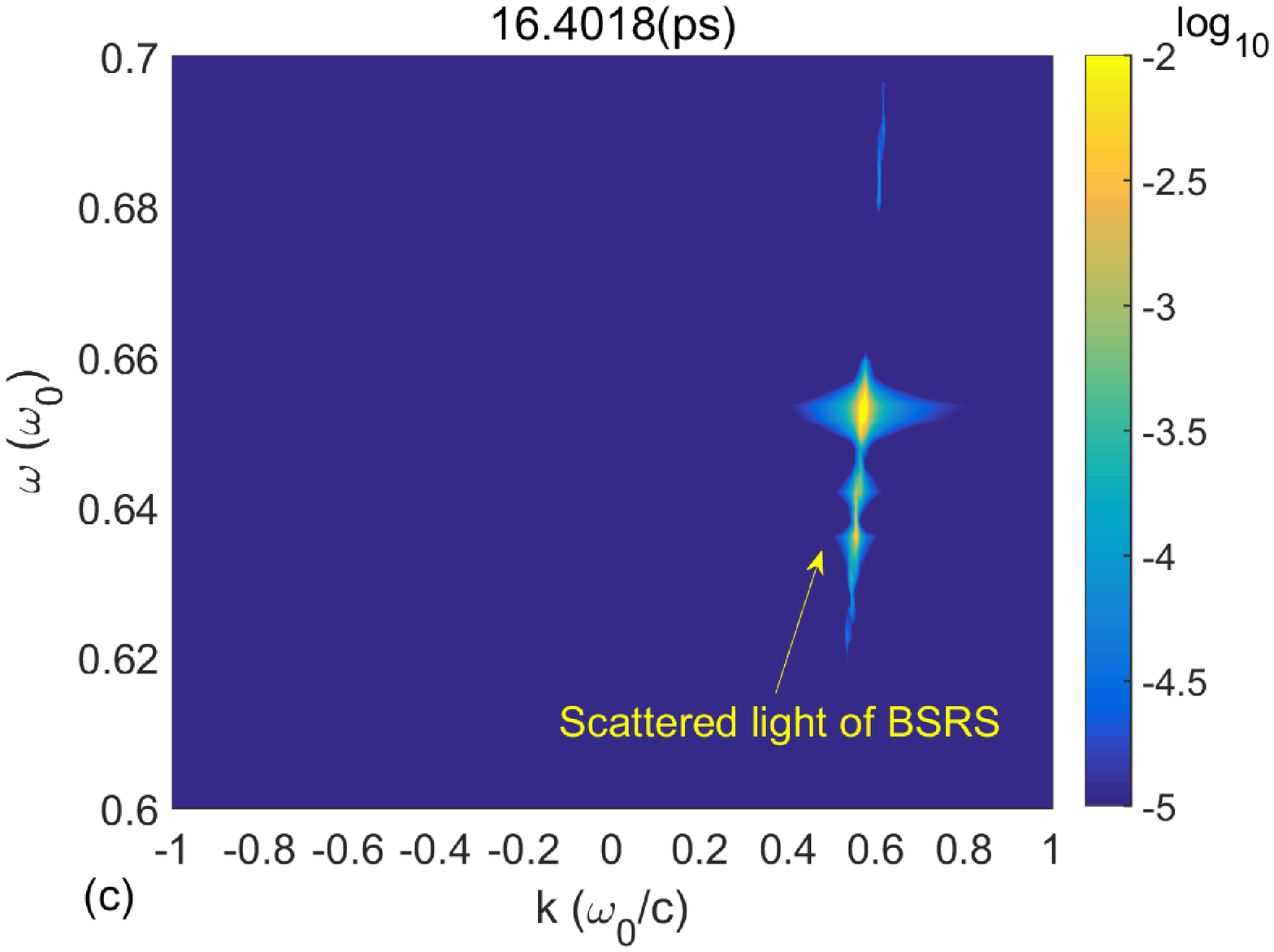}
\includegraphics[height=0.15\textwidth,width=0.23\textwidth,angle=0]{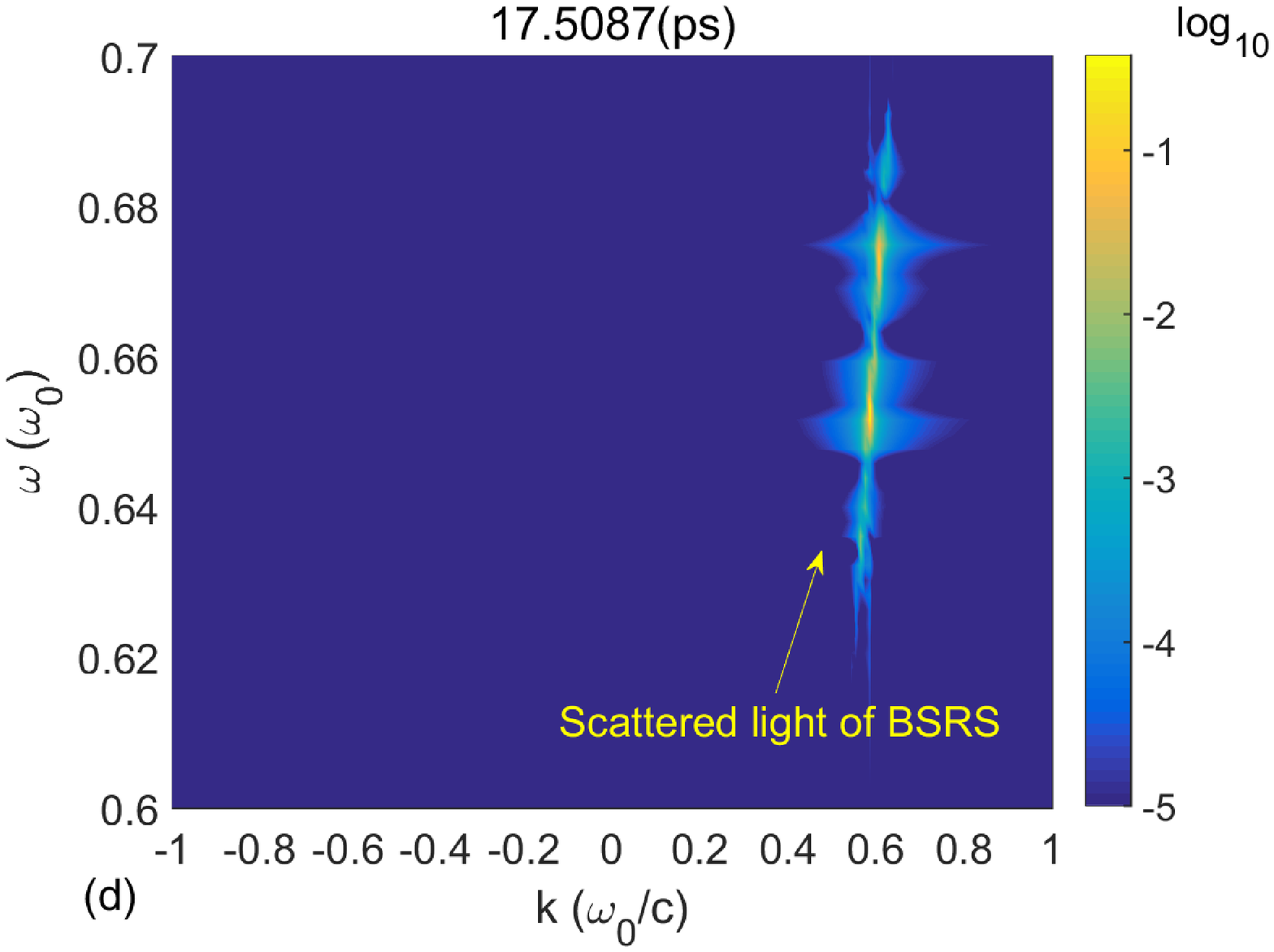}
\caption{(Color online) Snapshots of scattered light of SRS at the
time about (a,c) $16.4$ps and (b,d) $17.5$ps in (a,b) real space and
(c,d) $\omega-k$ space for the mobile ions case.}
\label{fig_ion_snapshots}
\end{figure}

In Fig. \ref{fig_ion_snapshots}(a), we show the scattered light of
BSRS at about $16.4$ ps. Similar to the fixed ions case shown in Fig
.\ref{fig_snapshots} (c), the scattered light of BSRS is strongly
depleted in the region of 8200 - 8800 $c/\omega_0$ due to re-SRS.
However, it is re-birthed at the left side of this region. Fig.
\ref{fig_ion_snapshots}(b) shows the scattered light of BSRS at a
later time 17.5 ps, in which the front part of the re-birthed
backscattered light is amplified convectively to a much higher level
than its saturated level before the growth of re-SRS as shown in
Fig. \ref{fig_snapshots}(a). We traced the convective inflated front
part of the re-birthed scattered light of BSRS, and gave its
$\omega-k$ spectra in the range of 7000 - 7600 $c/\omega_0$ and 600
- 1200 $c/\omega_0$ at 16.4 ps and 17.5 ps in Figs.
\ref{fig_ion_snapshots}(c) and \ref{fig_ion_snapshots}(d),
respectively. Compared with the single mode in Fig.
\ref{fig_psnapshots}(a), some modes of frequencies higher than 0.64
$\omega_0$, which corresponds to the best matching modes of BSRS at
different low density points, become stronger and stronger along
with the backward propagation. Based on these features, the reason
for the inflated BSRS is concluded as follow. First, re-SRS is
stimulated by the low frequency modes of backscattered light
originated from the high density region and induces LDI and even
cascade LDI in the region of 8200 - 8800 $c/\omega_0$. Second, the
low frequency density modulations of LDIs generate some modes of
scattered light of BSRS in a wider range of frequency locally in
this special region. These modes could even be the local absolute
modes when laser intensity was enough higher than the threshold
\cite{Lijun2017}. Third, different frequencies of these modes could
fulfill the best matching condition of BSRS at different low density
points as they propagate to the low density region. So these modes
finally work as a band of seeds and resulted in the amplification of
the engine modes for different low density points together on the
ray path.

In the experiments with gas-filled hohlraums at NIF, the wavelength
of scattered light of BSRS is generally shorter than the simulation
which is based on the linear ray-tracing model
\cite{Hall2017,Strozzi2017}. Here we present an explicit explanation
by investigating the coupling BSRS and re-SRS. Commonly, electron
density grows like a continuous exponential function of distance
from low density to a density higher than $n_c$ on the ray path in
hohlraums. In linear theory, low frequency (long wavelength) modes
of scattered light of BSRS matched to a high density point (above
about $0.1n_c$) should exist and even be stronger. Because the
higher density, the higher growth rate and spatial gain for SRS, as
long as the high density points exists in the gas region. However,
these lower-frequency modes can nevertheless meet their effective
quarter critical density points as they propagates to the low
density region and have chance to be heavily depleted by re-SRS.
While the modes originated from the density points lower than
$0.1n_c$ commonly have the higher frequencies, of which the
effective quarter critical density points are higher than $0.1n_c$.
As a result, these high-frequency (short wavelength) modes have no
chance to meet their effective quarter critical density again when
propagating to the much lower density region, and would be remained
or even be enhanced by re-SRS.

Furthermore, re-SRS would transfer partial energy from BSRS
scattered light into its daughter waves. This transferred energy can
be deposited quickly in the low density gas plasmas, and has almost
no contribution to the conversion into soft X rays. Due to the tiny
reflectivity, the rescattered light of re-SRS is hard to be observed
directly in the experiments and can not be considered in the
hydrodynamic simulation of NIF hohlraums. As a result, this would be
a potential reason for the ``drive deficit'' problem
\cite{MacLaren2014}. Experimentally, Thomson Scattering can be used
to diagnose the daughter waves of re-SRS \cite{Glenzer1996}.

In summary, coupling evolutions of SRS and re-SRS are studied in the
indirect drive regime with FLAME code for the first time. Re-SRS is
effectively found to be a frequency filter of BSRS light, which
heavily depletes the low frequency modes originated from high
density points and inflates the high frequency modes generated from
low density points, in the typical parameter space which is relevant
to the gas-filled hohlraum experiments at NIF. The simulation
results reveal the reason why the experimental measured wavelength
of BSRS light is always shorter than the simulated results based on
the linear theory, and also partially explain the ``energy deficit"
appeared in the gas-filled hohlraum experiments at NIF.

%\section{acknowledgments}
This work was supported by Science Challenge Project (Grant No.
TZ2016005) and the National Natural Science Foundation of China
(Grant No. 11875093 and No. 11775033).

%\section{REFERENCES}

\end{document}